\documentclass[aps,prd,showpacs,nofootinbib,tightenlines]{revtex4}
\usepackage{amsmath}
\usepackage{amssymb}
\usepackage{epsfig}
\usepackage{graphicx}
\usepackage{hyperref}
\usepackage{color}

\begin{document}

\title{Quasi-two-body decays  $B_{(s)} \to P D_0^*(2400) \to P D \pi$ in the perturbative QCD approach}
\author{Bo-Yan Cui$^{1,2}$}\email{boyancui@outlook.com}
\author{Ying-Ying Fan$^3$}\email{fyy163@126.com}
\author{Fu-Hu Liu$^{1,2}$}\email{fuhuliu@sxu.edu.cn}
\author{Wen-Fei Wang$^{1,2}$}\email{wfwang@sxu.edu.cn}
\affiliation{$^{1}$Institute of Theoretical Physics, Shanxi University, Taiyuan, Shanxi 030006, China}
\affiliation{$^{2}$State Key Laboratory of Quantum Optics and Quantum Optics Devices, Shanxi University, Taiyuan, Shanxi 030006, China}
\affiliation{$^{3}$College of Physics and Electronic Engineering, Xinyang Normal University, Xinyang 464000, China}
\date{\today}

\begin{abstract}
We study the quasi-two-body decays $B\to P D^{\ast}_0(2400)  \to P D\pi $ with $P=(\pi, K, \eta, \eta^{\prime})$ in the
perturbative QCD factorization approach. The predicted branching fractions for the considered decays are in the range
of $10^{-9}$-$10^{-4}$.
The strong Cabibbo-Kobayashi-Maskawa (CKM) suppression factor $R_{CKM}\approx \lambda^4 (\bar{\rho}^2 + \bar{\eta}^2) \approx 3\times 10^{-4}$
results in the great difference of the branching ratios for the decays with $D_0^*$ and $\bar{D}_0^*$ as the intermediate
states. The ratio $R_{\bar{D}_0^{*0}}$ between the decays $B^0 \to \bar{D}_0^{*0} K^0\to D^-\pi^+K^0$ and
$B^0 \to \bar{D}_0^{*0}\pi^0 \to D^-\pi^+\pi^0$ is about  $0.091^{+0.003}_{-0.005}$, consistent with the flavour-$SU$(3) symmetry result.
The ratio for the branching fractions is found to be $1.10^{+0.05}_{-0.02}$ between $\mathcal{B}(B_s^0\to D_0^{*+}K^-\to D^0\pi^+K^-)$
and $\mathcal{B}(B^0\to D_0^{*+} \pi^-\to D^0\pi^+ \pi^-)$ and to be $1.03^{+0.06}_{-0.07}$ between
$\mathcal{B}(B_s^0\to\bar{D}_0^{*0} \bar{K}^0\to D^-\pi^+ \bar{K}^0)$ and
$2\mathcal{B}(B^0\to \bar{D}_0^{*0}\pi^0\to D^-\pi^+\pi^0)$.
The predictions in this work can be tested by the future experiments.
\end{abstract}

\pacs{13.20.He, 13.25.Hw, 13.30.Eg}
\maketitle

\section{INTRODUCTION}
The strong dynamics contained in the three-body hadronic
$B$ meson decays is much more complicated than that in the two-body cases.
There are resonant and nonresonant contributions, final-state interactions~\cite{Bediaga:2015mia,Bediaga:2017axw},
and complex interplay between the weak processes and the low-energy strong interactions~\cite{Charles:2017ptc}
in the three-body $B$ meson decays. The traditional approaches
for the two-body decays are no longer satisfactory in the three-body processes~\cite{Amato:2016xjv}.
In order to extract the most information from the experimental data of those three-body processes, different methods
have been adopted abundantly in theoretical works~\cite{Wang:2018dfq}.
Three-body hadronic $B$ decays are known, in most cases, to be dominated by the low-energy scalar, vector,
and tensor resonant states.  In this situation, for the numerous three-body $B$ meson processes,
it is urgent to study the resonance contributions, which could be handled in the quasi-two-body framework where
the factorization procedure can be applied~\cite{Amato:2016xjv,Boito:2017jav}.

The $p$-wave orbitally excited state $D_0^*$~\footnote{For the sake of convenience, we employ $D_0^*$ to denote
$D_0^*(2400)$ in this work.}, with its $j_q=1/2$~\cite{Datta:2003re,Godfrey:2005ww,Godfrey:2015dva} and
$J^P=0^+$~\cite{Tanabashi:2018oca}, decays rapidly through $S$-wave pion emission.
It was thought to be the $c\bar{q}$ state in the traditional quark model~\cite{Godfrey:1985xj,Godfrey:1986wj,DiPierro:2001dwf},
but the mass observed in experiments~\cite{Abe:2003zm,Aubert:2009wg} is lower than the quark model predictions.
One possible explanation is that the self-energy hadronic loop could pull down the mass of the heavy scalar~\cite{Guo:2007up}
supported by~\cite{Cheng:2017oqh} within the framework of heavy meson chiral perturbation theory.
The tetraquark structure for $D_0^*$ was investigated in~\cite{Bracco:2005kt} with the help of the QCD sum rule,
and the authors of~\cite{Bracco:2005kt} suggested that the charmed scalar meson $D_0^{*0}(2308)$ observed by the
Belle Collaboration~\cite{Abe:2003zm} and $D_0^{*0(+)}(2405)$ observed by the FOCUS Collaboration~\cite{Link:2003bd} are different resonances.
It was claimed that two poles exist in the $D_0^*$ energy region~\cite{Albaladejo:2016lbb}, which has been supported by the lattice QCD analysis~\cite{Moir:2016srx}.
The resonant state $D_0^*$ has also been explained as a mixture of $c\bar{q}$ and tetraquarks~\cite{Vijande:2006hj} or a meson-meson bound state~\cite{Gamermann:2006nm}.
Since the Belle Collaboration's announcement~\cite{Abe:2003zm}, much work~\cite{Cheng:2003sm,Jugeau:2005yr,Cheng:2003id,Cheng:2006dm,Chen:2003rt} has emerged for the two-body hadronic $B$ decays involving $D_0^*$.

\begin{figure}[tbp]
\centerline{\epsfxsize=13cm \epsffile{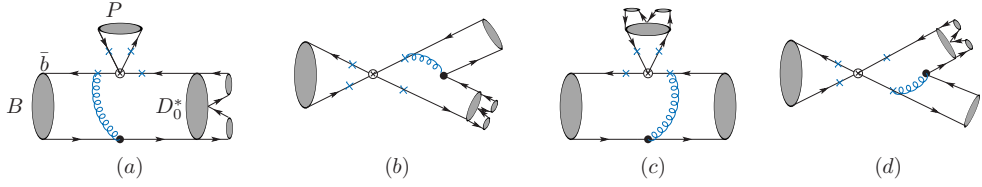}}
\vspace{0.3cm}
\caption{Typical diagrams for the quasi-two-body decays $B_{(s)}\to PD^{*}_{0}\to PD\pi$. The diagram (a) for the $B\to D_0^*$ transition, and diagram (c) for the $B\to P$ transition, as well as the diagrams (b) and (d) for for annihilation contributions. The symbol $\otimes$ stands
for the weak vertex and $\times$ denotes possible attachments of hard gluons.}
\label{fig-fig1}
\end{figure}

By studying the three-body hadronic $B$ meson decays involving $D_0^*$, one could provide the constraint on the unitary
triangle~\cite{Aaij:2016bqv,Gershon:2009qc,Craik:2017dpc,Bondar:2018gpb} and probe the inner structure of the intermediate
resonances. In Ref~\cite{Wang:2018fai}, four quasi-two-body decay processes involving $D_0^*$ have been studied
in the perturbative QCD (PQCD) approach~\cite{Keum:2000ph,Keum:2000wi,Lu:2000em,Li:2003yj}. In this work,
we extend the study to the quasi-two-body decays $B_{(s)} \to P D_0^* \to P D \pi$, with the bachelor particle $P$
which denotes the light pseudoscalar $\pi$, $K$, $\eta$, or $\eta^{\prime}$.
Typical diagrams for the $B_{(s)}\to PD^{*}_{0}\to PD\pi$ decays' processes are shown in Fig.~\ref{fig-fig1}.
Inspired by the generalized parton distribution in hard exclusive
two pion production~\cite{Diehl:1998dk,Mueller:1998fv,Polyakov:1998ze,Hagler:2002nh},
the two-meson distribution amplitude was introduced in three-body hadronic $B$ decays in~\cite{Chen:2002th,Chen:2004az}
as the universal nonperturbative input within the PQCD approach.
The PQCD approach has been employed in~\cite{Chen:2002th,Chen:2004az,Wang:2014ira,Wang:2015uea,Wang:2017hao}
 for the three-body
and in~\cite{Wang:2016rlo,Li:2016tpn,Li:2017mao,Li:2018qrm,Ma:2016csn,Ma:2017kec} for the quasi-two-body $B$ meson decays.
The decay amplitude for a three-body or quasi-two-body $B$ decay can be expressed as the convolution of the
nonperturbative wave function and hard kernel~\cite{Chen:2002th,Chen:2004az,Wang:2016rlo}. Taking
$B\to P D_0^*\to P D \pi$ as an example, we have the decay amplitude
\begin{eqnarray}
\mathcal{A}=\phi_B \otimes H \otimes \phi_P \otimes \phi_{D\pi}^{\text{S-wave}},
\end{eqnarray}
where hard kernel $H$ is calculated at leading order which contains one hard gluon, and the distribution amplitudes
$\phi_B, \phi_P$ and $\phi_{D\pi}^{\text{S-wave}}$ absorb  the nonperturbative dynamics in the decay processes.

The layout of this paper is as follows. We give a brief introduction of the theoretical framework in Sec.~\ref{section2}. Then the numerical results, a discussion and conclusions are given in Sec.~\ref{section3} and \ref{section4}. The relevant factorization formulas for the decay amplitudes are collected in the Appendix. 

\section{FRAMEWORK\label{section2}}

The definitions of the momenta for the $B_{(s)}$ meson, $S$-wave $D\pi$ system, and the bachelor meson are the same as those
in Ref.~\cite{Wang:2018fai}. The distribution amplitude and the parameters for the $S$-wave $D\pi$
system employed in this work as the same as those in~\cite{Wang:2018fai}.
The wave functions for $B_{(s)}$ and the relevant parameters can be found in~\cite{Wang:2012ab}.
The decay constants $f_{B^{0,\pm}}=0.190$ GeV for $B^{0,\pm}$ and $f_{B_{s}^{0}}=0.230$ GeV for $B_{s}^{0}$ were adopted
from recent lattice QCD updated results with $N_f=2+1+1$~\cite{Aoki:2019cca}.
The physical states $\eta$ and $\eta^{\prime}$
are related to the flavor states $\eta_q$ and $\eta_s$ via~\cite{Thomas:2007uy,Feldmann:1998vh,Feldmann:1998sh}
\begin{eqnarray}
\left(\begin{array}{c}\vert \eta \rangle \\ \vert \eta^{\prime} \rangle \end{array}\right)=
\left(\begin{array}{cc} \text{cos}\phi & -\text{sin}\phi \\ \text{sin}\phi & \text{cos}\phi \end{array} \right)
\left(\begin{array}{c}\vert \eta_q \rangle \\ \vert \eta_s  \rangle \end{array}\right)\;,
\end{eqnarray}
with the decay constants $f_q=(1.07\pm0.02)f_{\pi}$ and $f_s=(1.34\pm0.06)f_{\pi}$ for $\eta_q$ and $\eta_s$, respectively,
and the mixing angle  $\phi=39.3^{\circ} \pm 1.0^{\circ}$, which is close to the recent measurement
$\phi=(40.1\pm1.4_{\mathrm{stat}}\pm0.5_{\mathrm{syst}})^{\circ}$ by the BES\uppercase\expandafter{\romannumeral3}
Collaboration~\cite{Ablikim:2019rjz}. The wave functions for the states $\pi, K, \eta_q$ and $\eta_s$ in this work are written as
\begin{eqnarray}
\Phi_{P}(p,z)=\frac{i}{\sqrt{2N_c}}\gamma_5\left[p\hspace{-1.6truemm}/ \ \phi^A(z)+m_0\phi^P(z)+
m_0(v\hspace{-1.5truemm}/ n\hspace{-1.8truemm}/-1)\phi^T(z)\right]\;,
\end{eqnarray}
where $m_0$ is the chiral mass, $n=(1,0,\mathbf{0}_T)$ and $v=(0,1,\mathbf{0}_T)$ are the dimensionless lightlike unit vectors, $p$ and $z$ are, respectively, the momentum and corresponding momentum fraction of states $\pi, K, \eta_q$, and $\eta_s$. The distribution amplitudes $\phi^A(z), \phi^P(z), \phi^T(z)$ can be written as~\cite{Ball:1998tj,Ball:1998je,Ball:2004ye,Ball:2006wn}
\begin{eqnarray}
\phi^A(z)&=&\frac{f_{P}}{2\sqrt{2N_c}}6z(1-z)\left[1+a_1^{P}C_1^{3/2}(2z-1)+
a_2^{P}C_2^{3/2}(2z-1)+a_4^{P}C_4^{3/2}(2z-1)\right]\;, \nonumber\\
\phi^P(z)&=&\frac{f_{P}}{2\sqrt{2N_c}}\bigg[1+(30\eta_3-\frac{5}{2}\rho^2_{P})C_2^{1/2}(2z-1)-
3[\eta_3\omega_3+\frac{9}{20}\rho^2_{P}(1+6a_2^{P})]C_4^{1/2}(2z-1)\bigg]\;, \nonumber\\
\phi^T(z)&=&\frac{f_{P}}{2\sqrt{2N_c}}(1-2z)\bigg[1+6\left(5\eta_3-\frac{1}{2}\eta_3\omega_3-\frac{7}{20}\rho^2_{P}-
\frac{3}{5}\rho^2_{P}a_2^{P}\right)(1-10z+10z^2)\bigg]\;,
\end{eqnarray}
where the Gegenbauer moments are $a_1^{\pi,\eta_{q,s}}=0,\ a_1^K=0.06,\ a_2^{\pi,K}=0.25,\ a_2^{\eta_{q,s}}=0.115,\ a_4^{\pi,\eta_{q,s}}=-0.015$, and the parameters are $\rho_{\pi}=m_{\pi}/m_0^{\pi},\ \rho_{K}=m_{K}/m_0^{K},\ \rho_{\eta_q}=2m_q/m_{qq},\ \rho_{\eta_s}=2m_s/m_{ss}$, $\eta_3=0.015,\ \omega_3=-3$. Where $m_q$ is the mass of the up or down quark, $m_s$ is the mass of the strange quark, $m_{qq,ss}$ are related to $m_0^{\eta_q,\eta_s}$ by $m_0^{\eta_q}=m^2_{qq}/(m_u+m_d)$ and $m_0^{\eta_s}=m^2_{ss}/2m_s$, respectively. We adopt
$m_0^{\pi}=(1.4 \pm 0.1)$ GeV, $m_0^{K}=(1.6 \pm 0.1)$ GeV, $m_0^{\eta_q}=1.07$ GeV, and $m_0^{\eta_s}=1.92$ GeV in the numerical calculation. The Gegenbauer polynomials are defined as
\begin{eqnarray}
C_1^{\frac{3}{2}}(t)=3t, \quad C_2^{\frac{1}{2}}(t)=\frac{1}{2}(3t^2-1), \quad C_2^{\frac{3}{2}}(t)=\frac{3}{2}(5t^2-1),\nonumber\\
C_4^{\frac{1}{2}}(t)=\frac{1}{8}(3-30t^2+35t^4), \quad C_4^{\frac{3}{2}}(t)=\frac{15}{8}(1-14t^2+21t^4),
\end{eqnarray}
where the variable $t=2z-1$.

\section{RESULTS\label{section3}}

For the numerical calculations, we adopt from~\cite{Tanabashi:2018oca} the masses and mean lifetimes for the $B^{0,\pm}$ and $B_s^0$ mesons, the pole masses
and width for $D_0^{*0,\pm}$, the masses and decay constants for the light pseudoscalar mesons pion and kaon, and the
Wolfenstein parameters as:
\begin{eqnarray}
m_{B^{\pm,0}}&=&5.279,\quad m_{B_{s}^{0}}=5.367,\quad \tau_{B^{0}}=1.520,\quad \tau_{B^{\pm}}=1.638,\quad
\tau_{B_{s}^{0}}=1.509, \nonumber\\
m_{D_0^{*0}}&=&2.318, \quad m_{D_0^{*\pm}}=2.351, \quad \Gamma_{D_0^{*0}}=0.267,\quad \Gamma_{D_0^{*\pm}}=0.230,
\quad m_{\pi^{0}}=0.135, \nonumber\\
m_{\pi^{\pm }}&=&0.140, \quad m_K=0.496,\quad m_{\eta}=0.548,\quad
m_{\eta^{\prime}}=0.958, \quad f_K=0.156, \nonumber\\
f_{\pi }&=&0.130,\quad A=0.836, \quad \lambda=0.22453,\quad \bar{\eta}=0.355, \quad
\bar{\rho}=0.122,
\end{eqnarray}
where the masses, decay constants and widths are in units of GeV and lifetimes in units of $ps$.

By using the decay amplitudes for the decays $B_{(s)} \to PD_0^*\to PD\pi$ in the Appendix
and the differential branching fraction ($\mathcal{B}$), Eq.~(13) in~\cite{Wang:2018fai},
we obtain the branching fractions for the decays involving $B^+$ in Table~\ref{tableB+},
the results for the processes including $B^0$ in Table~\ref{tableB0}, and the values for the $B_s^0$ decay
modes in Table~\ref{tableBs} with the existing data from~\cite{Abe:2003zm,Aubert:2009wg,Aaij:2016fma,Aaij:2015vea,Kuzmin:2006mw,Aaij:2015sqa,Aaij:2015kqa}.
The first error of these results in Tables \ref{tableB0}-\ref{tableBs}  comes from the shape parameters
$\omega_{B^{0,\pm}}=0.40 \pm 0.04$ GeV for $B^{0,\pm}$ and $\omega_{B_s^0}=0.5 \pm 0.05$ GeV  for
$B_s^0$~\cite{Wang:2012ab}.
The second error comes from the shape parameter $\omega_{D\pi}=0.40 \pm 0.10$ GeV for the $D\pi$ system,
and the Gegenbauer moment $a_{D\pi}=0.40 \pm 0.10$ produces the third one~\cite{Wang:2018fai}.
The last one comes from the uncertainty of
decay width $\Gamma_{D_0^{*0}}=267 \pm 40$ MeV or $\Gamma_{D_0^{*+}}=230 \pm 17$ MeV~\cite{Tanabashi:2018oca}.
We have neglected the errors induced by the uncertainties of the parameters in the distribution amplitudes of the light pseudoscalar mesons and the Wolfenstein parameters since they are very small.

\begin{table}[thb]
\begin{center}
\caption{PQCD predictions for branching fractions of the quasi-two-body decays $B^+\to D_0^*P\to D\pi P$ together with the available experimental data.}
\footnotesize{
\begin{tabular}{l c c l}
  \hline\hline
  \     ~~~~~~~~Mode       & Unit & $\mathcal{B}$  &   ~~~~~~~~~~~~~~~~Data  \\  \hline
  $B^+\to D_0^{*0} \pi^+ \to D^+\pi^- \pi^+$           ~&~ ($10^{-8}$) ~&~ $1.13^{+0.36}_{-0.26}(\omega_{B}) ^{+0.13}_{-0.14}(\omega_{D\pi}) ^{+0.03}_{-0.05}(a_{D\pi})^{+0.06}_{-0.05}(\Gamma_{D_0^{*0}})$ ~&~  -  \\
  $B^+\to \bar{D}_0^{*0} \pi^+ \to D^-\pi^+ \pi^+$     ~&~ ($10^{-4}$) ~&~$5.95^{+2.37}_{-1.64}(\omega_{B}) ^{+1.97}_{-1.55}(\omega_{D\pi}) ^{+0.54}_{-0.49}(a_{D\pi})^{+0.29}_{-0.21}(\Gamma_{D_0^{*0}})$ ~&~  RPP\cite{Tanabashi:2018oca}: $6.4\pm 1.4$  \\
  $ $ ~&~ $ $ ~&~ $ $ ~&~ Belle\cite{Abe:2003zm}: $6.1\pm0.6\pm0.9\pm1.6 $ \\
  $ $ ~&~ $ $ ~&~ $ $ ~&~ BaBar\cite{Aubert:2009wg}: $6.8\pm0.3\pm0.4\pm2.0 $ \\
  $ $ ~&~ $ $ ~&~ $ $ ~&~ LHCb\cite{Aaij:2016fma}: $5.78\pm0.08\pm0.06\pm0.09\pm0.39$ \\
  $B^+\to D_0^{*0} K^+ \to D^+\pi^- K^+$               ~&~ ($10^{-7}$) ~&~ $3.56^{+1.02}_{-0.78}(\omega_{B}) ^{+0.46}_{-0.52}(\omega_{D\pi}) ^{+0.09}_{-0.15}(a_{D\pi})^{+0.16}_{-0.12}(\Gamma_{D_0^{*0}})$ ~&~  -   \\
  $B^+\to \bar{D}_0^{*0} K^+ \to D^-\pi^+ K^+$         ~&~ ($10^{-5}$) ~&~ $4.65^{+1.89}_{-1.30}(\omega_{B}) ^{+1.51}_{-1.24}(\omega_{D\pi}) ^{+0.40}_{-0.38}(a_{D\pi})^{+0.22}_{-0.18}(\Gamma_{D_0^{*0}})$ ~&~ LHCb\cite{Aaij:2015vea}: $0.61\pm0.19\pm0.05\pm0.14\pm0.04 $  \\
  $B^+\to D_0^{*+} \pi^0 \to D^0\pi^+ \pi^0$           ~&~ ($10^{-7}$) ~&~ $1.40^{+0.48}_{-0.34}(\omega_{B})
  ^{+0.02}_{-0.01}(\omega_{D\pi})^{+0.01}_{-0.00} (a_{D\pi})^{+0.03}_{-0.02}(\Gamma_{D_0^{*+}})$ ~&~ -  \\
  $B^+\to D_0^{*+} K^0 \to D^0\pi^+ K^0$               ~&~ ($10^{-9}$) ~&~ $5.52^{+0.15}_{-0.21}(\omega_{B}) ^{+1.73}_{-1.42}(\omega_{D\pi}) ^{+0.41}_{-0.36}(a_{D\pi})^{+0.13}_{-0.12}(\Gamma_{D_0^{*+}})$ ~&~ -  \\
  $B^+\to D_0^{*+} \eta \to D^0\pi^+ \eta$             ~&~ ($10^{-8}$) ~&~ $6.26^{+2.11}_{-1.49}(\omega_{B}) ^{+0.04}_{-0.03}(\omega_{D\pi}) ^{+0.03}_{-0.02}(a_{D\pi})^{+0.14}_{-0.10}(\Gamma_{D_0^{*+}})$ ~&~ -  \\
  $B^+\to D_0^{*+}\eta^{\prime}\to D^0\pi^+\eta^{\prime}$ ~&~ ($10^{-8}$) ~&~ $4.01^{+1.34}_{-0.96}(\omega_{B})
  ^{+0.02}_{-0.03}(\omega_{D\pi})^{+0.02}_{-0.01}(a_{D\pi})^{+0.07}_{-0.06}(\Gamma_{D_0^{*+}})$ ~&~ -  \\
  \hline\hline
  \label{tableB+}
\end{tabular}}
\vspace{-0.5cm}
\end{center}
\end{table}
\begin{table}[thb]
\begin{center}
\caption{PQCD prediction of branching fraction for the quasi-two-body decays $B^0\to D_0^*P\to D\pi P$ together with the available experimental data.}
\footnotesize{
\begin{tabular}{l c c l}
  \hline\hline
  \     ~~~~~~~~Mode       & Unit & $\mathcal{B}$  &  ~~~~~~~~~~~~~~~~Data   \\  \hline
  $B^0\to D_0^{*-} \pi^+ \to \bar{D}^0\pi^- \pi^+$              ~&~ ($10^{-4}$) ~&~ $2.85^{+1.23}_{-0.80}(\omega_{B}) ^{+1.05}_{-0.81}(\omega_{D\pi}) ^{+0.33}_{-0.31}(a_{D\pi})^{+0.06}_{-0.05}(\Gamma_{D_0^{*+}})$ ~&~ RPP\cite{Tanabashi:2018oca}: $0.76\pm 0.08$  \\
    $ $ ~&~ $ $ ~&~ $ $ ~&~ Belle\cite{Kuzmin:2006mw}: $0.60\pm0.13\pm0.15\pm0.22 $ \\
  $ $ ~&~ $ $ ~&~ $ $ ~&~ LHCb\cite{Aaij:2015sqa}: $0.77\pm0.05\pm0.03\pm0.03\pm0.04\footnotemark[1] $ \\
  $ $ ~&~ $ $ ~&~ $ $ ~&~ LHCb\cite{Aaij:2015sqa}: $0.80\pm0.05\pm0.08\pm0.04\pm0.04\footnotemark[2] $ \\
  $B^0\to D_0^{*+} \pi^- \to D^0\pi^+ \pi^-$                    ~&~ ($10^{-7}$) ~&~ $2.56^{+0.85}_{-0.65}(\omega_{B}) ^{+0.01}_{-0.02}(\omega_{D\pi}) ^{+0.02}_{-0.03}(a_{D\pi})^{+0.03}_{-0.06}(\Gamma_{D_0^{*+}})$  ~&~ -  \\
  $B^0\to D_0^{*-} K^+ \to \bar{D}^0\pi^- K^+$                  ~&~ ($10^{-5}$) ~&~ $2.38^{+0.95}_{-0.65}(\omega_{B}) ^{+0.85}_{-0.68}(\omega_{D\pi}) ^{+0.30}_{-0.28}(a_{D\pi})^{+0.04}_{-0.03}(\Gamma_{D_0^{*+}})$  ~&~ LHCb\cite{Aaij:2015kqa}: $1.77\pm0.26\pm0.19\pm0.67\pm0.20$    \\
  $B^0\to D_0^{*0} \pi^0 \to D^+\pi^- \pi^0$                    ~&~ ($10^{-9}$) ~&~ $4.20^{+1.62}_{-1.07}(\omega_{B}) ^{+0.44}_{-0.48}(\omega_{D\pi}) ^{+0.09}_{-0.07}(a_{D\pi})^{+0.07}_{-0.12}(\Gamma_{D_0^{*0}})$  ~&~  - \\
  $B^0\to \bar{D}_0^{*0} \pi^0 \to D^-\pi^+ \pi^0$              ~&~ ($10^{-5}$) ~&~ $2.29^{+0.87}_{-0.61}(\omega_{B}) ^{+0.51}_{-0.43}(\omega_{D\pi}) ^{+0.09}_{-0.06}(a_{D\pi})^{+0.12}_{-0.04}(\Gamma_{D_0^{*0}})$  ~&~ -\\
  $B^0\to D_0^{*0} K^0 \to D^+\pi^- K^0$                        ~&~ ($10^{-7}$) ~&~ $2.69^{+0.91}_{-0.66}(\omega_{B}) ^{+0.30}_{-0.32}(\omega_{D\pi}) ^{+0.09}_{-0.08}(a_{D\pi})^{+0.12}_{-0.11}(\Gamma_{D_0^{*0}})$  ~&~  - \\
  $B^0\to \bar{D}_0^{*0} K^0 \to D^-\pi^+ K^0$                  ~&~ ($10^{-6}$) ~&~ $4.15^{+1.54}_{-1.09}(\omega_{B}) ^{+0.74}_{-0.72}(\omega_{D\pi}) ^{+0.03}_{-0.03}(a_{D\pi})^{+0.19}_{-0.14}(\Gamma_{D_0^{*0}})$  ~&~  -  \\
  $B^0\to D_0^{*0} \eta \to D^+\pi^- \eta$                      ~&~ ($10^{-9}$) ~&~ $2.81^{+0.78}_{-0.58}(\omega_{B}) ^{+0.30}_{-0.33}(\omega_{D\pi}) ^{+0.11}_{-0.14}(a_{D\pi})^{+0.13}_{-0.09}(\Gamma_{D_0^{*0}})$  ~&~ - \\
  $B^0\to D_0^{*0} \eta^{\prime} \to D^+\pi^- \eta^{\prime}$    ~&~ ($10^{-9}$) ~&~ $1.80^{+0.49}_{-0.37}(\omega_{B}) ^{+0.19}_{-0.21}(\omega_{D\pi}) ^{+0.07}_{-0.09}(a_{D\pi})^{+0.08}_{-0.06}(\Gamma_{D_0^{*0}})$  ~&~  - \\
  $B^0\to \bar{D}_0^{*0} \eta \to D^-\pi^+ \eta$                ~&~ ($10^{-5}$) ~&~ $1.79^{+0.60}_{-0.41}(\omega_{B}) ^{+0.30}_{-0.28}(\omega_{D\pi}) ^{+0.07}_{-0.03}(a_{D\pi})^{+0.09}_{-0.06}(\Gamma_{D_0^{*0}})$  ~&~  - \\
  $B^0\to\bar{D}_0^{*0}\eta^{\prime}\to D^-\pi^+\eta^{\prime}$  ~&~ ($10^{-5}$) ~&~ $1.15^{+0.38}_{-0.27}(\omega_{B}) ^{+0.19}_{-0.18}(\omega_{D\pi}) ^{+0.04}_{-0.02}(a_{D\pi})^{+0.06}_{-0.04}(\Gamma_{D_0^{*0}})$  ~&~ - \\
  \hline\hline
  \label{tableB0}
\end{tabular}}
\footnotetext[1]{ Isobar model }
\footnotetext[2]{ K-matrix model }
\end{center}
\end{table}
\begin{table}[thb]
\begin{center}
\caption{PQCD prediction of branching fraction for the quasi-two-body decays $B^0_s\to D_0^*P\to D\pi P$.}
\begin{tabular}{l c c}
  \hline\hline
  \     ~~~~Mode       & Unit &  $\mathcal{B}$     \\  \hline
  $B^0_s\to D_0^{*-} \pi^+ \to \bar{D}^0\pi^- \pi^+$          ~&~ ($10^{-7}$) ~&~ $2.70^{+0.29}_{-0.36}(\omega_{B}) ^{+0.60}_{-0.58}(\omega_{D\pi}) ^{+0.43}_{-0.31}(a_{D\pi})^{+0.06}_{-0.01}(\Gamma_{D_0^{*+}})$   \\
  $B^0_s\to D_0^{*+} \pi^- \to D^0\pi^+ \pi^-$                ~&~ ($10^{-9}$) ~&~ $2.90^{+0.08}_{-0.15}(\omega_{B}) ^{+0.95}_{-0.83}(\omega_{D\pi}) ^{+0.26}_{-0.23}(a_{D\pi})^{+0.07}_{-0.06}(\Gamma_{D_0^{*+}})$    \\
  $B^0_s\to D_0^{*+} K^- \to D^0\pi^+ K^-$                    ~&~ ($10^{-7}$) ~&~ $2.82^{+1.09}_{-0.74}(\omega_{B}) ^{+0.02}_{-0.01}(\omega_{D\pi})^{+0.01}_{-0.00} (a_{D\pi})^{+0.06}_{-0.04}(\Gamma_{D_0^{*+}})$    \\
  $B^0_s\to D_0^{*0} \pi^0 \to D^+\pi^- \pi^0$                ~&~ ($10^{-9}$) ~&~ $1.48^{+0.03}_{-0.04}(\omega_{B}) ^{+0.46}_{-0.42}(\omega_{D\pi}) ^{+0.12}_{-0.13}(a_{D\pi})^{+0.08}_{-0.07}(\Gamma_{D_0^{*0}})$    \\
  $B^0_s\to \bar{D}_0^{*0} \pi^0 \to D^-\pi^+ \pi^0$          ~&~ ($10^{-7}$) ~&~ $1.38^{+0.24}_{-0.19}(\omega_{B}) ^{+0.48}_{-0.33}(\omega_{D\pi}) ^{+0.22}_{-0.16}(a_{D\pi})^{+0.07}_{-0.04}(\Gamma_{D_0^{*0}})$    \\
  $B^0_s\to D_0^{*0} \bar{K}^0 \to D^+\pi^- \bar{K}^0$        ~&~ ($10^{-9}$) ~&~ $9.09^{+3.65}_{-2.38}(\omega_{B}) ^{+0.84}_{-0.95}(\omega_{D\pi}) ^{+0.38}_{-0.23}(a_{D\pi})^{+0.41}_{-0.31}(\Gamma_{D_0^{*0}})$  \\
  $B^0_s\to \bar{D}_0^{*0} \bar{K}^0 \to D^-\pi^+ \bar{K}^0$  ~&~ ($10^{-5}$) ~&~ $4.70^{+2.05}_{-1.39}(\omega_{B}) ^{+0.76}_{-0.75}(\omega_{D\pi}) ^{+0.04}_{-0.05}(a_{D\pi})^{+0.21}_{-0.16}(\Gamma_{D_0^{*0}})$    \\
  $B^0_s\to D_0^{*0} \eta \to D^+\pi^- \eta$                  ~&~ ($10^{-8}$) ~&~ $9.37^{+4.31}_{-2.70}(\omega_{B}) ^{+0.67}_{-0.77}(\omega_{D\pi}) ^{+0.21}_{-0.15}(a_{D\pi})^{+0.43}_{-0.30}(\Gamma_{D_0^{*0}})$     \\
  $B^0_s\to D_0^{*0}\eta^{\prime}\to D^+\pi^-\eta^{\prime}$   ~&~ ($10^{-7}$) ~&~ $1.62^{+0.65}_{-0.43}(\omega_{B}) ^{+0.16}_{-0.15}(\omega_{D\pi}) ^{+0.06}_{-0.05}(a_{D\pi})^{+0.09}_{-0.05}(\Gamma_{D_0^{*0}})$    \\
  $B^0_s\to \bar{D}_0^{*0} \eta \to D^-\pi^+ \eta$            ~&~ ($10^{-6}$) ~&~ $1.27^{+0.55}_{-0.39}(\omega_{B}) ^{+0.18}_{-0.20}(\omega_{D\pi}) ^{+0.04}_{-0.03}(a_{D\pi})^{+0.05}_{-0.04}(\Gamma_{D_0^{*0}})$    \\
  $B^0_s\to\bar{D}_0^{*0}\eta^{\prime}\to D^-\pi^+\eta^{\prime}$ ~&~ ($10^{-6}$) ~&~ $2.24^{+0.93}_{-0.64}(\omega_{B}) ^{+0.29}_{-0.30}(\omega_{D\pi}) ^{+0.02}_{-0.04}(a_{D\pi})^{+0.11}_{-0.08}(\Gamma_{D_0^{*0}})$    \\
  \hline\hline
  \label{tableBs}
\end{tabular}
\end{center}
\end{table}


\begin{figure}[tbp]
\centerline{\epsfxsize=6cm \epsffile{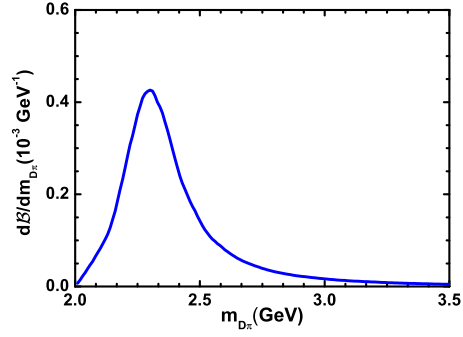}}
\vspace{0.3cm}
\caption{The $D\pi$ invariant mass-dependent differential branching fraction for $B^0\to\bar{D}_0^{*0}\pi^0\to D^-\pi^+\pi^0$.}
\label{fig-wdep}
\end{figure}

The four quasi-two-body decays $B^+\to \bar{D}_0^{*0} \pi^+ \to D^-\pi^+ \pi^+$,
$B^+\to \bar{D}_0^{*0} K^+ \to D^-\pi^+ K^+$, $B^0\to D_0^{*-} \pi^+ \to \bar{D}^0\pi^- \pi^+$ and
$B^0\to D_0^{*-} K^+ \to \bar{D}^0\pi^- K^+$ have been discussed in Ref.~\cite{Wang:2018fai}.
For completeness, we keep their branching ratios in Tables~\ref{tableB+} and \ref{tableB0}.
In Fig.~\ref{fig-wdep}, we show the $D\pi$ invariant mass-dependent differential branching fraction for the quasi-two-body decay $B^0\to\bar{D}_0^{*0}\pi^0\to D^-\pi^+\pi^0$.
One can find that the main portion of branching fraction for $B^0\to\bar{D}_0^{*0}\pi^0\to D^-\pi^+\pi^0$
comes from the region around the pole mass of the resonant state $D_0^*$. The contributions from the $m_{D\pi}$ mass region larger than $3$ GeV can be neglected safely as argued in Ref.~\cite{Wang:2018fai}.

For the CKM suppressed decay modes $B\to D_0^*\pi\to D\pi\pi$ and $B_s\to D_0^*\bar{K}\to D\pi\bar{K}$, their branching ratios
are much smaller than the corresponding results of $B\to \bar{D}_0^*\pi \to D\pi\pi$ and $B_s\to \bar{D}_0^*\bar{K} \to D\pi\bar{K}$
decays as predicted by PQCD in this work. The major reason comes from the strong CKM suppression factor~\cite{Ma:2016csn}
\begin{eqnarray}
R_{CKM}=\bigg|\frac{V_{ub}^*V_{cd}}{V_{cb}^*V_{ud}}\bigg|^2 \approx \lambda^4 (\bar{\rho}^2 + \bar{\eta}^2) \approx 3\times 10^{-4} \;.
\end{eqnarray}
For the CKM suppressed and CKM favored decay modes concerned in this work,
we define the following ratios of the branching fractions for the the corresponding decays as
\begin{eqnarray}\label{eq:Rckm}
R_1 &=&\frac{B^0\to D_0^{*0}\pi^0\to D^+\pi^-\pi^0}{B^0\to\bar{D}_0^{*0}\pi^0\to D^-\pi^+\pi^0}\approx 1.83 \times 10^{-4} \;,\nonumber\\
R_2 &=&\frac{B^0\to D_0^{*0}\eta\to D^+\pi^-\eta}{B^0\to\bar{D}_0^{*0}\eta\to D^-\pi^+\eta}\approx 1.57 \times 10^{-4} \;,\nonumber\\
R_3 &=&\frac{B^0\to D_0^{*0}\eta^{\prime}\to D^+\pi^-\eta^{\prime}}{B^0\to\bar{D}_0^{*0}\eta^{\prime}\to D^-\pi^+\eta^{\prime}} \approx 1.57 \times 10^{-4} \;, \nonumber\\
R_4 &=&\frac{B_s\to D_0^{*0}\bar{K}^0\to D^+\pi^-\bar{K}^0}{B_s\to\bar{D}_0^{*0}\bar{K}^0\to D^-\pi^+\bar{K}^0}\approx 1.93 \times 10^{-4}\;,\nonumber\\
R_5 &=&\frac{B^+\to D_0^{*0}\pi^+\to D^+\pi^-\pi^+}{B^+\to \bar{D}_0^{*0}\pi^+\to D^-\pi^+\pi^+}\approx 1.91 \times 10^{-5}\;.
\end{eqnarray}
The ratios $R_{1}, R_{2}, R_{3}$, and $R_{4}$ are close to each other, because all four decay pairs in these four ratios decay through the same colour suppressed emission topologies, and the nonfactorizable diagrams in Fig.~\ref{fig-fig1} play the dominant role.
The nonvanishing charm quark mass in the fermion propagator generates the main differences between the $R_{CKM}$ and
$R_{1,2,3,4}$. For the decay process $B^+\to\bar{D}_0^{*0}\pi^+\to D^-\pi^+\pi^+$, one has the contributions from  both the
$B\to \bar D_0^{*0}$ transition and the $B\to\pi$ transition, while for $B^+\to D_0^{*0}\pi^+\to D^+\pi^-\pi^+$, one has only the
colour suppressed transition $B\to\pi$. So it is not surprising to have a quite small value for $R_5$.

Assuming factorization and flavour-$SU$(3) symmetry, the ratio between the two decays $B^0\to D_0^{*-} K^+ \to \bar{D}^0\pi^- K^+$
and $B^0\to D_0^{*-} \pi^+ \to \bar{D}^0\pi^- \pi^+$ will not very far from 0.076, as discussed in Ref.~\cite{Wang:2018fai}.
The same situation should happen to the decays $B^0 \to \bar{D}_0^{*0} K^0\to D^-\pi^+K^0$ and
$B^0 \to \bar{D}_0^{*0}\pi^0 \to D^-\pi^+\pi^0$. With the PQCD predictions in Table~\ref{tableB0}, we have
\begin{eqnarray}
R_{\bar{D}_0^{*0}} &=& \frac{\mathcal{B}(B^0 \to \bar{D}_0^{*0} K^0\to D^-\pi^+K^0)}
{2\mathcal{B}(B^0 \to \bar{D}_0^{*0}\pi^0 \to D^-\pi^+\pi^0)}=0.091^{+0.003}_{-0.005}\;.
\end{eqnarray}
The deviation between the $R_{\bar{D}_0^{*0}}$ and
\begin{eqnarray}
\left|\frac{V_{us}}{V_{ud}}\right|^2\cdot \frac{f^2_K}{f^2_\pi}=0.076
\end{eqnarray}
could be due to the violation of the flavour-$SU$(3) symmetry and the contributions from annihilation diagrams in the
$B^0 \to \bar{D}_0^{*0}\pi^0 \to D^-\pi^+\pi^0$ process.

The ratio of branching fractions with topologically similar decay processes $B_s^0\to D_0^{*+}K^-\to D^0\pi^+K^-$ and
$B^0\to D_0^{*+} \pi^-\to D^0\pi^+ \pi^-$ is expected to be close to 1 in the na\"ive factorization because of the close
values for the $B_s^0\to K^-$ and $B^0 \to \pi^-$ transition form factors~\cite{Wang:2012ab}.
With the predictions in Tables~\ref{tableB0} and \ref{tableBs}, we have
\begin{eqnarray}
\frac{\mathcal{B} (B_s^0\to D_0^{*+}K^-\to D^0\pi^+K^-)}
{\mathcal{B}(B^0\to D_0^{*+} \pi^-\to D^0\pi^+ \pi^-)}=1.10^{+0.05}_{-0.02}\;.
\end{eqnarray}
A similar relation for $B_s^0\to\bar{D}_0^{*0} \bar{K}^0\to D^-\pi^+ \bar{K}^0$ and
$B^0\to \bar{D}_0^{*0}\pi^0\to D^-\pi^+\pi^0$ is
\begin{eqnarray}
\frac{\mathcal{B}(B_s^0\to\bar{D}_0^{*0} \bar{K}^0\to D^-\pi^+ \bar{K}^0)}
{2\mathcal{B}(B^0\to \bar{D}_0^{*0}\pi^0\to D^-\pi^+\pi^0)}=1.03^{+0.06}_{-0.07}\;
\end{eqnarray}
induced from Tables~\ref{tableB0} and \ref{tableBs}.

\section{CONCLUSION\label{section4}}
We have studied the quasi-two-body decays $B_{(s)} \to P D_0^* \to P D \pi$, where the bachelor particle $P$
denotes $\pi$, $K$, $\eta$, or $\eta^{\prime}$ in the PQCD approach.
The predicted branching fractions for the considered decays are in the range of $10^{-9}-10^{-4}$.
For the decays $B\to D_0^*\pi\to D\pi\pi$ and $B\to \bar{D}_0^*\pi\to D\pi\pi$ as well as $B_s\to D_0^*\to D\pi\bar{K}$ and
$B_s\to \bar{D}_0^*\bar{K}\to D\pi\bar{K}$, the great difference in their corresponding branching fractions can be
understood by a strong CKM suppression factor $R_{CKM}\approx \lambda^4 (\bar{\rho}^2 + \bar{\eta}^2) \approx 3\times 10^{-4}$.
The flavour-$SU$(3) symmetry can be employed to analyse the quasi-two-body decays with the same topologies, such as
$B^0 \to \bar{D}_0^{*0} K^0\to D^-\pi^+K^0$ and $B^0 \to \bar{D}_0^{*0}\pi^0 \to D^-\pi^+\pi^0$, while $R_{\bar{D}_0^{*0}}$ was
predicted to be $0.091^{+0.003}_{-0.005}$ for their branching ratios.  The ratio for the branching fractions was found to be $1.10^{+0.05}_{-0.02}$ between
$\mathcal{B}(B_s^0\to D_0^{*+}K^-\to D^0\pi^+K^-)$ and $\mathcal{B}(B^0\to D_0^{*+} \pi^-\to D^0\pi^+ \pi^-)$
and to be $1.03^{+0.06}_{-0.07}$ between $\mathcal{B}(B_s^0\to\bar{D}_0^{*0} \bar{K}^0\to D^-\pi^+ \bar{K}^0)$ and
$2\mathcal{B}(B^0\to \bar{D}_0^{*0}\pi^0\to D^-\pi^+\pi^0)$, which can
be tested by the precise data from the future experiments.

\acknowledgments
We are grateful to Muhammad Waqas for helpful comments. This work is supported in part by the National Natural Science Foundation of China under Grants No. 11547038, No. 11505148,  and No. 11575103.

\appendix*
\section{Decay Amplitudes\label{appendix}}
The amplitudes from Fig.~\ref{fig-fig1} are written as
\begin{eqnarray}
{\mathcal A}\big(B^+\to \pi^+[D_0^{*0}\to]D^+\pi^-\big)&=&\frac{G_F}{\sqrt2}V^*_{ub}V_{cd}
\bigg\{a_2F_{TP}+C_2 M^{\prime}_{TP}+a_1 F_{AD}+C_1M_{AD}\bigg\}\;,
\nonumber\\
\label{eqn=B+pi+D0}
{\mathcal A}\big(B^+\to \pi^+[\bar{D}_0^{*0}\to]D^-\pi^+\big)&=&\frac{G_F}{\sqrt2}V^*_{cb}V_{ud}
\bigg\{a_2F_{TP}+C_2 M_{TP}+a_1 F_{TD}+C_1M_{TD}\bigg\}\;,
\nonumber\\
\label{eqn=B+pi+D0bar}
{\mathcal A}\big(B^+\to K^+[D_0^{*0}\to]D^+\pi^-\big)&=&\frac{G_F}{\sqrt2}V^*_{ub}V_{cs}
\bigg\{a_2F_{TP}+C_2 M^{\prime}_{TP}+a_1 F_{AD}+C_1M_{AD}\bigg\}\;,
\nonumber\\
\label{eqn=B+K+D0}
{\mathcal A}\big(B^+\to K^+[\bar{D}_0^{*0}\to]D^-\pi^+\big)&=&\frac{G_F}{\sqrt2}V^*_{cb}V_{us}
\bigg\{a_2F_{TP}+C_2 M_{TP}+a_1 F_{TD}+C_1M_{TD}\bigg\}\;,
\nonumber\\
\label{eqn=B+K+D0bar}
{\mathcal A}\big(B^+\to \pi^0[D_0^{*+}\to]D^0\pi^+\big)&=&\frac{G_F}{2}V^*_{ub}V_{cd}
\bigg\{a_1 \left(F_{TP}-F_{AD}\right)+C_1 \left(M^{\prime}_{TP}-M_{AD}\right)\bigg\}\;,
\nonumber\\
\label{eqn=B+pi0D+}
{\mathcal A}\big(B^+\to K^0[D_0^{*+}\to]D^0\pi^+\big)&=&\frac{G_F}{\sqrt2}V^*_{ub}V_{cs}
\bigg\{a_1 F_{AD}+C_1 M_{AD}\bigg\}\;,\label{eqn=B+K0D+}\nonumber\\
{\mathcal A}\big(B^+\to \eta_q[D_0^{*+}\to]D^0\pi^+\big)&=&\frac{G_F}{2}V^*_{ub}V_{cd}
\bigg\{a_1 (F_{TP}+F_{AD})+C_1 (M^{\prime}_{TP}+M_{AD})\bigg\}\;,
\nonumber\\
\label{eqn=B+etaqD+}
{\mathcal A}\big(B^+\to \eta[D_0^{*+}\to]D^0\pi^+\big)&=&{\mathcal A}\big(B^+\to \eta_q[D_0^{*+}\to]D^0\pi^+\big) \cos\phi\;,
\nonumber\\
{\mathcal A}\big(B^+\to \eta^{\prime}[D_0^{*+}\to]D^0\pi^+\big)&=&{\mathcal A}\big(B^+\to \eta_q[D_0^{*+}\to]D^0\pi^+\big) \sin\phi\;,
\nonumber\\
{\mathcal A}\big(B^0\to \pi^+[D_0^{*-}\to]\bar{D}^0\pi^-\big)&=&\frac{G_F}{\sqrt2}V^*_{cb}V_{ud}
\bigg\{a_2 F_{AP}+C_2 M_{AP}+a_1 F_{TD}+C_1M_{TD}\bigg\}\;,
\nonumber\\
\label{eqn=B0pi+D-}
{\mathcal A}\big(B^0\to \pi^-[D_0^{*+}\to]D^0\pi^+\big)&=&\frac{G_F}{\sqrt2}V^*_{ub}V_{cd}
\bigg\{a_2 F_{AD}+C_2 M_{AD}+a_1 F_{TP}+C_1M^{\prime}_{TP}\bigg\}\;,
\nonumber
\label{eqn=B0pi-D+}
\end{eqnarray}
\begin{eqnarray}
{\mathcal A}\big(B^0\to K^+[D_0^{*-}\to]\bar{D}^0\pi^-\big)&=&\frac{G_F}{\sqrt2}V^*_{cb}V_{us}
\bigg\{a_1 F_{TD}+C_1 M_{TD}\bigg\}\;,
\nonumber\\
\label{eqn=B0 K+D-}
{\mathcal A}\big(B^0\to \pi^0[D_0^{*0}\to]D^+\pi^-\big)&=&\frac{G_F}{2}V^*_{ub}V_{cd}
\bigg\{a_2(F_{AD}-F_{TP})+C_2 \left(M_{AD}-M^{\prime}_{TP}\right)\bigg\}\;,
\nonumber\\
\label{eqn=B0pi0D0}
{\mathcal A}\big(B^0\to \pi^0[\bar{D}_0^{*0}\to]D^-\pi^+\big)&=&\frac{G_F}{2}V^*_{cb}V_{ud}
\bigg\{a_2(F_{AP}-F_{TP})+C_2 \left(M_{AP}-M_{TP}\right)\bigg\}\;,
\nonumber\\
\label{eqn=B0pi0D0bar}
{\mathcal A}\big(B^0\to K^0[D_0^{*0}\to]D^+\pi^-\big)&=&\frac{G_F}{\sqrt2}V^*_{ub}V_{cs}
\bigg\{a_2F_{TP}+C_2 M^{\prime}_{TP}\bigg\}\;,
\nonumber\\
\label{eqn=B0K0D0}
{\mathcal A}\big(B^0\to K^0[\bar{D}_0^{*0}\to]D^-\pi^+\big)&=&\frac{G_F}{\sqrt2}V^*_{cb}V_{us}
\bigg\{a_2F_{TP}+C_2 M_{TP}\bigg\}\;,
\nonumber\\
\label{eqn=B0K0D0bar}
{\mathcal A}\big(B^0\to \eta_q[D_0^{*0}\to]D^+\pi^-\big)&=&\frac{G_F}{2}V^*_{ub}V_{cd}
\bigg\{a_2(F_{TP}+F_{AD})+C_2(M^{\prime}_{TP}+M_{AD})\bigg\}\;,
\nonumber\\
\label{eqn=B0etaqD0}
{\mathcal A}\big(B^0\to \eta[D_0^{*0}\to]D^+\pi^-\big)&=&{\mathcal A}\big(B^0\to \eta_q[D_0^{*0}\to]D^+\pi^-\big) \cos\phi\;,
\nonumber\\
{\mathcal A}\big(B^0\to \eta^{\prime}[D_0^{*0}\to]D^+\pi^-\big)&=&{\mathcal A}\big(B^0\to \eta_q[D_0^{*0}\to]D^+\pi^-\big) \sin\phi\;,
\nonumber\\
{\mathcal A}\big(B^0\to \eta_q[\bar{D}_0^{*0}\to]D^-\pi^+\big)&=&\frac{G_F}{2}V^*_{cb}V_{ud}
\bigg\{a_2(F_{TP}+F_{AP})+C_2(M_{TP}+M_{AP})\bigg\}\;,
\nonumber\\
\label{eqn=B0etaqD0bar}
{\mathcal A}\big(B^0\to \eta[\bar{D}_0^{*0}\to]D^-\pi^+\big)&=&{\mathcal A}\big(B^0\to \eta_q[\bar{D}_0^{*0}\to]D^-\pi^+\big)\cos\phi\;,
\nonumber\\
{\mathcal A}\big(B^0\to \eta^{\prime}[\bar{D}_0^{*0}\to]D^-\pi^+\big)&=&{\mathcal A}\big(B^0\to \eta_q[\bar{D}_0^{*0}\to]D^-\pi^+\big) \sin\phi\;,
\nonumber\\
{\mathcal A}\big(B_s^0\to \pi^+[D_0^{*-}\to]\bar{D}^0\pi^-\big)&=&\frac{G_F}{\sqrt2}V^*_{cb}V_{us}
\bigg\{a_2 F_{AP}+C_2 M_{AP}\bigg\}\;,
\nonumber\\
\label{eqn=Bspi+D-}
{\mathcal A}\big(B_s^0\to \pi^-[D_0^{*+}\to]D^0\pi^+\big)&=&\frac{G_F}{\sqrt2}V^*_{ub}V_{cs}
\bigg\{a_2 F_{AD}+C_2 M_{AD}\bigg\}\;,
\nonumber\\
\label{eqn=Bspi-D+}
{\mathcal A}\big(B_s^0\to K^-[D_0^{*+}\to]D^0\pi^+\big)&=&\frac{G_F}{\sqrt2}V^*_{ub}V_{cd}
\bigg\{a_1 F_{TP}+C_1 M^{\prime}_{TP}\bigg\}\;,
\nonumber\\
\label{eqn=BsK-D+}
{\mathcal A}\big(B_s^0\to \pi^0[D_0^{*0}\to]D^+\pi^-\big)&=&\frac{G_F}{2}V^*_{ub}V_{cs}
\bigg\{a_2 F_{AD}+C_2 M_{AD}\bigg\}\;,
\nonumber\\
\label{eqn=Bspi0D0}
{\mathcal A}\big(B_s^0\to \pi^0[\bar{D}_0^{*0}\to]D^-\pi^+\big)&=&\frac{G_F}{2}V^*_{cb}V_{us}
\bigg\{a_2 F_{AP}+C_2 M_{AP}\bigg\}\;,
\nonumber\\
\label{eqn=Bspi0D0bar}
{\mathcal A}\big(B_s^0\to \bar{K}^0[D_0^{*0}\to]D^+\pi^-\big)&=&\frac{G_F}{\sqrt2}V^*_{ub}V_{cd}
\bigg\{a_2F_{TP}+C_2 M^{\prime}_{TP}\bigg\}\;,
\nonumber\\
\label{eqn=BsK0barD0}
{\mathcal A}\big(B_s^0\to \bar{K}^0[\bar{D}_0^{*0}\to]D^-\pi^+\big)&=&\frac{G_F}{\sqrt2}V^*_{cb}V_{ud}
\bigg\{a_2F_{TP}+C_2 M_{TP}\bigg\}\;,
\nonumber\\
\label{eqn=BsK0bar0D0bar}
{\mathcal A}\big(B_s^0\to \eta_q[D_0^{*0}\to]D^+\pi^-\big)&=&\frac{G_F}{2}V^*_{ub}V_{cs}
\bigg\{a_2F_{AD}+C_2 M_{AD}\bigg\}\;,
\nonumber\\
\label{eqn=BsetaqD0}
{\mathcal A}\big(B_s^0\to \eta_s[D_0^{*0}\to]D^+\pi^-\big)&=&\frac{G_F}{\sqrt2}V^*_{ub}V_{cs}
\bigg\{a_2F_{TP}+C_2 M^{\prime}_{TP}\bigg\}\;,
\nonumber\\
\label{eqn=BsetasD0}
{\mathcal A}\big(B_s^0\to \eta[D_0^{*0}\to]D^+\pi^-\big)&=&{\mathcal A}\big(B_s^0\to \eta_q[D_0^{*0}\to]D^+\pi^-\big)\cos\phi -
{\mathcal A}\big(B_s^0\to \eta_s[D_0^{*0}\to]D^+\pi^-\big) \sin\phi\;,
\nonumber\\
{\mathcal A}\big(B_s^0\to \eta^{\prime}[D_0^{*0}\to]D^+\pi^-\big)&=&{\mathcal A}\big(B_s^0\to \eta_q[D_0^{*0}\to]D^+\pi^-\big)\sin\phi +
{\mathcal A}\big(B_s^0\to \eta_s[D_0^{*0}\to]D^+\pi^-\big) \cos\phi\;,
\nonumber\\
{\mathcal A}\big(B_s^0\to \eta_q[\bar{D}_0^{*0}\to]D^-\pi^+\big)&=&\frac{G_F}{2}V^*_{cb}V_{us}
\bigg\{a_2F_{AP}+C_2 M_{AP}\bigg\}\;,
\nonumber\\
\label{eqn=BsetaqD0bar}
{\mathcal A}\big(B_s^0\to \eta_s[\bar{D}_0^{*0}\to]D^-\pi^+\big)&=&\frac{G_F}{\sqrt2}V^*_{cb}V_{us}
\bigg\{a_2F_{TP}+C_2 M_{TP}\bigg\}\;,
\nonumber\\
\label{eqn=BsetasD0bar}
{\mathcal A}\big(B_s^0\to \eta[\bar{D}_0^{*0}\to]D^-\pi^+\big)&=&{\mathcal A}\big(B_s^0\to\eta_q[\bar{D}_0^{*0}\to]D^-\pi^+\big)\cos\phi-
{\mathcal A}\big(B_s^0\to \eta_s[\bar{D}_0^{*0}\to]D^-\pi^+\big)\sin\phi\;,
\nonumber\\
{\mathcal A}\big(B_s^0\to \eta^{\prime}[\bar{D}_0^{*0}\to]D^-\pi^+\big)&=&{\mathcal A}\big(B_s^0\to\eta_q[\bar{D}_0^{*0}\to]D^-\pi^+\big)\sin\phi+
{\mathcal A}\big(B_s^0\to \eta_s[\bar{D}_0^{*0}\to]D^-\pi^+\big)\cos\phi\;,\nonumber
\end{eqnarray}
where $G_F$ is the Fermi constant, $V$'s are the CKM matrix elements, $C_1$ and $C_2$ are Wilson coefficients
and $a_1=C_1/3+C_2$ and $a_2=C_2/{3}+C_1$. The factorization formulas for decay amplitudes from Fig.~\ref{fig-fig1} are collected below:
\begin{eqnarray}
F_{TD} &=& 8\pi C_F m^4_B f_P\int dx_B dx_3\int b_B db_B b_3 db_3 \phi_B(x_B,b_B)\phi_{D\pi}(x_3,b_3,s)(\eta-1)\big\{\big[\sqrt{\eta}(2x_3-1)-x_3-1 \big]
\nonumber\\
&\times& E_{1ab}(t_{1a})h_{1a}(x_B,x_3,b_B,b_3)+\left(\eta+2\sqrt{\eta}(r_c-1)-r_c \right)E_{1ab}(t_{1b})h_{1b}(x_B,x_3,b_B,b_3) \big\}\;, \nonumber\\
\label{flltd}
M_{TD} &=& 32\pi C_F m^4_B/\sqrt{2N_c} \int dx_B dz dx_3\int b_B db_B b db\phi_B(x_B,b_B)\phi_{D\pi}(x_3,b_3,s)\phi^A(\eta-1)
\nonumber\\
&\times&\big\{\left[\eta\left(1-x_3-z\right)+z+x_B+x_3\sqrt{\eta}-1\right]E_{1cd}(t_{1c})h_{1c}(x_B,z,x_3,b_B,b)
\nonumber\\
&+&\left[z\left(1-\eta\right)-x_B+x_3\left(1-\sqrt{\eta}\right) \right]E_{1cd}(t_{1d})h_{1d}(x_B,z,x_3,b_B,b) \big\}\;, \nonumber\\
\label{mlltd}
F_{AD} &=& 8\pi C_F m^4_B f_B\int dz dx_3\int b db b_3 db_3\phi_{D\pi}(x_3,b_3,s)\big\{\left[\phi^A(\eta-1)(1-x_3)-2\phi^P\left(x_3-2\right) \sqrt{\eta}r_0\right] E_{1ef}(t_{1e})
\nonumber\\
&\times& h_{1e}(z,x_3,b,b_3)+\big[(\eta-1)[2r_c\sqrt{\eta}+z(\eta-1)-\eta]\phi^A+2r_0\sqrt{\eta}(\eta-1)[z(\phi^P+\phi^T)-\phi^T]
\nonumber\\
&+&r_0(\eta+1)(-2\sqrt{\eta}+r_c)\phi^P+r_0r_c(\eta-1)\phi^T \big] \times E_{1ef}(t_{1f})h_{1f}(z,x_3,b,b_3) \big\}\;,\nonumber\\
\label{fllad}
M_{AD} &=& 32\pi C_F m^4_B/\sqrt{2N_c} \int dx_B dz dx_3\int b_B db_B b db\phi_B(x_B,b_B)\phi_{D\pi}(x_3,b_3,s)\big\{\big[(1-\eta)z[(\eta-1)\phi^A
\nonumber\\
   &+&\sqrt{\eta}r_0(\phi^P+\phi^T)]+x_B[(\eta-1)\phi^A+\sqrt{\eta}r_0(\phi^P+\phi^T)]+\sqrt{\eta}[\eta r_0(\phi^P+\phi^T)-r_0(x_3-3)\phi^P
\nonumber\\
   &+&r_0(x_3-1)\phi^T+\sqrt{\eta}(\eta-1)(1-x_3)\phi^A]\big]\times E_{1gh}(t_{1g})h_{1g}(x_B,z,x_3,b_B,b)
\nonumber\\
   &+&\big[\sqrt{\eta}r_0[(\phi^P-\phi^T)(\eta z-\eta-z+x_B)+(\phi^P+\phi^T)(x_3-1)]
\nonumber\\
   &+&\phi^A(\eta^2-1)(x_3-1)\big]E_{1gh}(t_{1h})h_{1h}(x_B,z,x_3,b_B,b)\big\}\;,\nonumber\\
\label{mllad}
F_{TP} &=& 8\pi C_F m^4_B F_{D\pi}(s) \int dx_B dz\int b_B db_B b db \phi_B(x_B,b_B)\big\{\big[\phi^A(1-\eta)(z(\eta-1)-1)
\nonumber\\
   &-&r_0[\phi^P(\eta+2(\eta-1)z+1)+\phi^T(\eta-1)(2z-1)]\big]E_{2ab}(t_{2a})h_{2a}(x_B,z,b_B,b)
\nonumber\\
   &+&\left[2r_0\phi^P(\eta+\eta x_B-1)+(\eta-1)\eta x_B\phi^A\right]E_{2ab}(t_{2b})\times h_{2b}(x_B,z,b_B,b)\big\}\;,\nonumber\\
\label{flltp}
M_{TP} &=& 32\pi C_F m^4_B/\sqrt{2N_c} \int dx_B dz dx_3\int b_B db_B b_3 db_3\phi_B(x_B,b_B)\phi_{D\pi}(x_3,b_3,s)
\big\{\big[(\eta-1)((\eta+1)(x_B+x_3-1)
\nonumber\\
&-&r_c\sqrt{\eta})\phi^A+r_0[z(1-\eta)(\phi^T-\phi^P) +(x_B+x_3)\eta(\phi^T+\phi^P)-(2\eta+4r_c\sqrt{\eta})\phi^P]\big]
\nonumber\\
&\times&E_{2cd}(t_{2c})h_{2c}(x_B,z,x_3,b_B,b_3)-\big[(\eta-1)z[(\eta-1)\phi^A+r_0(\phi^P+\phi^T)]
\nonumber\\
&+&(x_B-x_3)[\eta r_0(\phi^P-\phi^T)+(\eta-1)\phi^A]\big] E_{2cd}(t_{2d})h_{2d}(x_B,z,x_3,b_B,b_3)\big\}\;, \nonumber\\
\label{mlltp}
M^{\prime}_{TP} &=& 32\pi C_F m^4_B/\sqrt{2N_c} \int dx_B dz dx_3\int b_B db_B b_3 db_3\phi_B(x_B,b_B)\phi_{D\pi}(x_3,b_3,s)
\big\{\big[(1-x_B-x_3)(1-\eta^2)\phi^A
\nonumber\\
&+&r_0[z(1-\eta)(\phi^T-\phi^P) +(x_B+x_3)\eta(\phi^T+\phi^P)-2\eta\phi^P]\big] E_{2cd}(t^{\prime}_{2c})h^{\prime}_{2c}(x_B,z,x_3,b_B,b_3)
\nonumber\\
&+&\big[(\eta-1)[(1-\eta)z-x_B-r_c\sqrt{\eta}+x_3]\phi^A-r_0z(\eta-1)(\phi^P+\phi^T)
\nonumber\\
&+&r_0\eta(x_B-x_3)(\phi^T-\phi^P)-4r_0r_c\sqrt{\eta}\phi^P\big]\times E_{2cd}(t^{\prime}_{2d})h^{\prime}_{2d}(x_B,z,x_3,b_B,b_3)\big\}\;,\nonumber\\
F_{AP} &=& 8\pi C_F m^4_B f_B \int dz dx_3\int b db b_3 db_3 \phi_{D\pi}(x_3,b_3,s) \big\{\big[(\eta-1)[2\sqrt{\eta}r_c+(\eta-1)z+1] \phi^A\nonumber\\
&-&r_0[(\eta+1)r_c+2\sqrt{\eta}(z(\eta-1)+2)]\phi^P+r_0(\eta-1)(r_c+2\sqrt{\eta}z)\phi^T \big]E_{2ef}(t_{2e})h_{2e}(z,x_3,b,b_3)
\nonumber\\
&+& \left[2\sqrt{\eta}r_0\phi^P(-\eta+x_3+1)-\phi^A(\eta-1)x_3\right]E_{2ef}(t_{2f})h_{2f}(z,x_3,b,b_3) \big\},\nonumber\\
M_{AP} &=& 32\pi C_F m^4_B/\sqrt{2N_c} \int dx_B dz dx_3\int b_B db_B b db\phi_B(x_B,b_B)\phi_{D\pi}(x_3,b_3,s) \big\{\big[\eta\phi^A(1-\eta)
\nonumber\\
  &+&\sqrt{\eta}[-\eta r_0(z-1)(\phi^P+\phi^T)+r_0(z-3)\phi^P+r_0(z-1)\phi^T]+(x_B+x_3)[(\eta^2-1)\phi^A
\nonumber\\
  &+&r_0\sqrt{\eta}(\phi^T-\phi^P)]\big]E_{2gh}(t_{2g})h_{2g}(x_B,z,x_3,b_B,b)+\big[(1-\eta)\phi^A[\eta(x_3-x_B+z-1)-z+1]
\nonumber\\
 &+&r_0\sqrt{\eta}[(\eta-1)(z-1)(\phi^P-\phi^T)+(x_3-x_B)(\phi^P+\phi^T)] \big]E_{2gh}(t_{2h})h_{2h}(x_B,z,x_3,b_B,b) \big\},
\nonumber
\end{eqnarray}
where $x_B$, $x_3$, and $z$ are momentum fractions of the corresponding spectator quarks, as defined in Ref.~\cite{Wang:2018fai}. $b_B$, $b_3$, and $b$ are the conjugate variables of transverse momenta $P_B$, $P_3$, and $P$, respectively. Variable $\eta$ is defined as $\eta=m^2_{D\pi}/m^2_B$. The ratio $r_0=m_0/m_B$, where $m_0$ is the chiral mass of light pseudoscalars. $r_c=m_c/m_B$ is the ratio of the charm quark mass to the $B$ meson mass. The functions $E_{1mn}$ and $E_{2mn}$($m=a,c,e,g$ and $n=b,d,f,h$) are the evolution factors, which are given by
\begin{eqnarray}
E_{1ab}(t)&=&\alpha(t)\exp[-S_B(t)-S_D(t)],  \qquad E_{2ab}(t)=\alpha(t)\exp[-S_B(t)-S_P(t)],  \nonumber\\
E_{1cd}(t)&=&\alpha(t)\exp[-S_B(t)-S_D(t)-S_P(t)]_{b_3=b_B}, \quad E_{2cd}(t)=\alpha(t)\exp[-S_B(t)-S_D(t)-S_P(t)]_{b=b_B},  \nonumber\\
E_{1ef}(t)&=&\alpha(t)\exp[-S_P(t)-S_D(t)],  \qquad E_{2ef}(t)=E_{1ef}(t),  \nonumber\\
E_{1gh}(t)&=&\alpha(t)\exp[-S_B(t)-S_D(t)-S_P(t)]_{b=b_3}, \quad E_{2gh}(t)=E_{1gh}(t),	\nonumber
\end{eqnarray}
in which Sudakov exponents $S_{(B,D,P)}(t)$ are defined as
\begin{eqnarray}
S_B(t)&=&s\big(\frac{x_B m_B}{\sqrt2},b_B\big)+2\int^{t}_{1/b_B}\frac{d\bar{\mu}}{\bar{\mu}}\gamma_q(\alpha_s(\bar{\mu}))\;, \nonumber\\
S_D(t)&=&s\big(\frac{x_3 m_B}{\sqrt2},b_3\big)+
       2\int^{t}_{1/b_3}\frac{d\bar{\mu}}{\bar{\mu}}\gamma_q(\alpha_s(\bar{\mu}))\;,\nonumber\\
S_P(t)&=&s\big(\frac{z m_B}{\sqrt2},b\big)+s\big(\frac{(1-z) m_B}{\sqrt2},b\big)+
       2\int^{t}_{1/b}\frac{d\bar{\mu}}{\bar{\mu}}\gamma_q(\alpha_s(\bar{\mu}))\;,\nonumber
\end{eqnarray}
where the quark anomalous dimension $\gamma_q =-\alpha_s/\pi$. The explicit form for $s(Q,b)$ at one loop can be found in~\cite{Ali:2007ff}. $t_{1x}$ and $t_{2x}$($x=a,b\cdots h$) are hard scales which are chosen to be the maximum of the virtuality of the internal momentum transition in the hard amplitudes as
\begin{eqnarray}
t_{1a}&=& Max \big\{m_B\sqrt{x_3},1/b_B,1/b_3\big\}\;,\nonumber\\
t_{1b}&=& Max \big\{m_B\sqrt{x_3 x_B},m_B\sqrt{\vert x_B-\eta+r_c^2 \vert},1/b_B,1/b_3\big\}\;,\nonumber\\
t_{1c}&=& Max \big\{m_B\sqrt{x_3 x_B},m_B\sqrt{\vert x_3[x_B-(1-\eta)(1-z)] \vert},1/b_B,1/b\big\}\;,\nonumber\\
t_{1d}&=& Max \big\{m_B\sqrt{x_3 x_B},m_B\sqrt{\vert x_3[x_B-(1-\eta)z] \vert},1/b_B,1/b\big\}\;,\nonumber\\
t_{1e}&=& Max \big\{m_B\sqrt{(1-x_3)[(1-\eta)z+\eta]},m_B\sqrt{1-x_3},1/b_3,1/b\big\}\;,\nonumber\\
t_{1f}&=& Max \big\{m_B\sqrt{(1-x_3)[(1-\eta)z+\eta]},m_B\sqrt{\vert \eta+(1-\eta)z-r_c^2 \vert},1/b_3,1/b\big\}\;,\nonumber\\
t_{1g}&=& Max \big\{m_B\sqrt{(1-x_3)[(1-\eta)z+\eta]},m_B\sqrt{1-x_3[(1-\eta)(1-z)-x_B]},1/b_B,1/b\big\}\;, \nonumber\\
t_{1h}&=& Max \big\{m_B\sqrt{(1-x_3)[(1-\eta)z+\eta]},m_B\sqrt{\vert (1-x_3)[x_B-\eta-(1-\eta)z] \vert},1/b_B,1/b\big\}\;, \nonumber\\
t_{2a}&=& Max \big\{m_B\sqrt{(1-\eta)z},1/b_B,1/b\big\}\;,\nonumber\\
t_{2b}&=& Max \big\{m_B\sqrt{(1-\eta)x_B},1/b_B,1/b\big\}\;,\nonumber\\
t_{2c}&=& Max \big\{m_B\sqrt{(1-\eta)z x_B},m_B\sqrt{\vert r_c^2-(1-x_3-x_B)[(1-\eta)z+\eta] \vert },1/b_B,1/b_3\big\}\;, \nonumber\\
t_{2d}&=& Max \big\{m_B\sqrt{(1-\eta)z x_B},m_B\sqrt{(1-\eta)z \vert x_B-x_3 \vert},1/b_B,1/b_3\big\}\;,\nonumber\\
t^{\prime}_{2c}&=& Max \big\{m_B\sqrt{(1-\eta)z x_B},m_B\sqrt{\vert 1-x_3-x_B \vert [(1-\eta)z+\eta]},1/b_B,1/b_3\big\}\;,\nonumber\\
t^{\prime}_{2d}&=& Max \big\{m_B\sqrt{(1-\eta)z x_B},m_B\sqrt{\vert r_c^2+(1-\eta)z(x_B-x_3) \vert},1/b_B,1/b_3\big\}\;,\nonumber\\
t_{2e}&=& Max \big\{m_B\sqrt{(1-\eta)(1-z)x_3},m_B\sqrt{\vert1-r_c^2-(1-\eta)z\vert},1/b_3,1/b\big\}\;,\nonumber\\
t_{2f}&=& Max \big\{m_B\sqrt{(1-\eta)x_3},1/b_3,1/b\big\}\;,\nonumber\\
t_{2g}&=& Max \big\{m_B\sqrt{(1-\eta)(1-z)x_3},m_B\sqrt{\vert [\eta+(1-\eta)z](1-x_3-x_B)-1 \vert},1/b_B,1/b\big\}\;,\nonumber\\
t_{2h}&=& Max \big\{m_B\sqrt{(1-\eta)(1-z)x_3},m_B\sqrt{\vert x_3-x_B \vert (1-\eta)(1-z)},1/b_B,1/b\big\}\;.\nonumber
\end{eqnarray}
The hard functions can be written as
\begin{eqnarray}
h_{1a}(x_B,x_3,b_B,b_3)&=&K_0(m_B\sqrt{x_3 x_B}b_B)[K_0(m_B\sqrt{x_3}b_B)I_0(m_B\sqrt{x_3}b_3)\theta(b_B-b_3)+ (b_B \longleftrightarrow b_3)]S_t(x_3), \nonumber\\
h_{1b}(x_B,x_3,b_B,b_3)&=&K_0(m_B\sqrt{x_3 x_B}b_3)S_t(x_B)   \nonumber\\
              &\times&\left\{ \begin{array}{ll} [\theta(b_3-b_B)K_0(m_B\sqrt{r_c^2+x_B-\eta}b_3) \\
              \times I_0(m_B\sqrt{r_c^2+x_B-\eta}b_B)+(b_3\longleftrightarrow b_B)],\quad &r_c^2+x_B \geq \eta, \\
              \frac{i\pi}{2}[\theta(b_3-b_B)H^{(1)}_0(m_B\sqrt{\eta-x_B-r_c^2}b_3) \\
              \times J_0(m_B\sqrt{\eta-x_B-r_c^2}b_B)+(b_3\longleftrightarrow b_B)],\quad &r_c^2+x_B < \eta,\end{array} \right.  \nonumber\\
h_{1c}(x_B,z,x_3,b_B,b)&=&[K_0(m_B\sqrt{x_3 x_B}b_B)I_0(m_B\sqrt{x_3 x_B}b)\theta(b_B-b)+(b \longleftrightarrow b_B) ] \nonumber\\
              &\times&\left\{ \begin{array}{ll} K_0(m_B\sqrt{x_3[x_B-(1-\eta)(1-z)]}b), \quad &x_B \geq (1-\eta)(1-z), \\
              \frac{i\pi}{2}H^{(1)}_0(m_B\sqrt{x_3[(1-\eta)(1-z)-x_B]}b), \quad &x_B < (1-\eta)(1-z),\end{array} \right.  \nonumber\\
h_{1d}(x_B,z,x_3,b_B,b)&=&[K_0(m_B\sqrt{x_3 x_B}b_B)I_0(m_B\sqrt{x_3 x_B}b)\theta(b_B-b)+(b \longleftrightarrow b_B) ] \nonumber\\
              &\times&\left\{ \begin{array}{ll} K_0(m_B\sqrt{x_3[x_B-(1-\eta)z]}b), \quad &x_B \geq (1-\eta)z, \\
              \frac{i\pi}{2}H^{(1)}_0(m_B\sqrt{x_3[(1-\eta)z-x_B]}b), \quad &x_B < (1-\eta)z,\end{array} \right.  \nonumber\\
h_{1e}(z,x_3,b,b_3)&=&(\frac{i\pi}{2})^2 H^{(1)}_0(m_B\sqrt{(1-x_3)[\eta+z(1-\eta)]}b)[H^{(1)}_0(m_B\sqrt{1-x_3}b) \nonumber\\
              &\times&J_0(m_B\sqrt{1-x_3}b_3)\theta(b-b_3)+(b_3 \longleftrightarrow b)]S_t(x_3), \nonumber\\
h_{1f}(z,x_3,b,b_3)&=&\frac{i\pi}{2} H^{(1)}_0(m_B\sqrt{(1-x_3)[\eta+z(1-\eta)]}b_3)S_t(z) \nonumber\\
                &\times&\left\{ \begin{array}{ll} [\theta(b_3-b)K_0(m_B\sqrt{r_c^2-[\eta+(1-\eta)z]}b_3) \\
                \times I_0(m_B\sqrt{r_c^2-[\eta+(1-\eta)z]}b)+(b \longleftrightarrow b_3) ]\quad &r_c^2 \geq \eta+(1-\eta)z, \\
              \frac{i\pi}{2}[\theta(b_3-b)H^{(1)}_0(m_B\sqrt{[\eta+(1-\eta)z]-r_c^2}b_3) \\
             \times J_0(m_B\sqrt{[\eta+(1-\eta)z]-r_c^2}b)+(b \longleftrightarrow b_3)] \quad &r_c^2
               < \eta+(1-\eta)z,\end{array} \right.  \nonumber\\
h_{1g}(x_B,z,x_3,b_B,b)&=&\frac{i\pi}{2} K_0(m_B\sqrt{1-x_3[(1-z)(1-\eta)-x_B]}b_B) [H^{(1)}_0(m_B\sqrt{(1-x_3)[\eta+z(1-\eta)]}b_B) \nonumber\\
              &\times&J_0(m_B\sqrt{(1-x_3)[\eta+z(1-\eta)]}b)\theta(b_B-b)+(b \longleftrightarrow b_B)], \nonumber\\
h_{1h}(x_B,z,x_3,b_B,b)&=&[\frac{i\pi}{2}H^{(1)}_0(m_B\sqrt{(1-x_3)[\eta+z(1-\eta)]}b_B)  \nonumber\\
              &\times&J_0(m_B\sqrt{(1-x_3)[\eta+z(1-\eta)]}b)\theta(b_B-b)+(b \longleftrightarrow b_B) ] \nonumber\\
              &\times&\left\{ \begin{array}{ll} K_0(m_B\sqrt{(1-x_3)[x_B-\eta-z(1-\eta)]}b_B), \quad &x_B \geq \eta+z(1-\eta), \\
              \frac{i\pi}{2}H^{(1)}_0(m_B\sqrt{(1-x_3)[-x_B+\eta+z(1-\eta)]}b_B), \quad &x_B < \eta+z(1-\eta),\end{array} \right.  \nonumber\\
h_{2a}(x_B,z,b_B,b)&=&K_0(m_B\sqrt{(1-\eta)z x_B}b_B)[K_0(m_B\sqrt{(1-\eta)z}b_B) \nonumber\\
              &\times&I_0(m_B\sqrt{(1-\eta)z}b)\theta(b_B-b)+(b \longleftrightarrow b_B)]S_t(z), \nonumber\\
h_{2b}(x_B,z,b_B,b)&=&K_0(m_B\sqrt{(1-\eta)z x_B}b)[K_0(m_B\sqrt{(1-\eta)x_B}b) \nonumber\\
              &\times&I_0(m_B\sqrt{(1-\eta)x_B}b_B)\theta(b-b_B)+(b \longleftrightarrow b_B)]S_t(x_B), \nonumber\\
h_{2c}(x_B,z,x_3,b_B,b_3)&=&[K_0(m_B\sqrt{(1-\eta)z x_B}b_B)I_0(m_B\sqrt{(1-\eta)z x_B}b_3)\theta(b_B-b_3) +(b_3 \longleftrightarrow b_B) ] \nonumber\\
              &\times&\left\{ \begin{array}{ll} K_0(m_B\sqrt{r_c^2-[\eta+(1-\eta)z](1-x_B-x_3)}b_3), \\ \quad r_c^2 \geq [\eta+(1-\eta)z](1-x_B-x_3), \\
              \frac{i\pi}{2}H^{(1)}_0(m_B\sqrt{[\eta+(1-\eta)z](1-x_B-x_3)-r^2_c}b_3), \\ \quad r^2_c < [\eta+(1-\eta)z](1-x_B-x_3),\end{array} \right.  \nonumber\\
h_{2d}(x_B,z,x_3,b_B,b_3)&=&[K_0(m_B\sqrt{(1-\eta)z x_B}b_B)I_0(m_B\sqrt{(1-\eta)z x_B}b_3)\theta(b_B-b_3) +(b_3 \longleftrightarrow b_B) ] \nonumber\\
              &\times&\left\{ \begin{array}{ll} K_0(m_B\sqrt{(1-\eta)(x_B-x_3)z}b_3), \quad &x_B \geq x_3,\\
              \frac{i\pi}{2}H^{(1)}_0(m_B\sqrt{(1-\eta)(x_3-x_B)z}b_3), \quad &x_B < x_3,\end{array} \right.  \nonumber
\end{eqnarray}
\begin{eqnarray}
h^{\prime}_{2c}(x_B,z,x_3,b_B,b_3)&=&[K_0(m_B\sqrt{(1-\eta)z x_B}b_B)I_0(m_B\sqrt{(1-\eta)z x_B}b)\theta(b_B-b_3)+(b_3 \longleftrightarrow b_B) ] \nonumber\\
              &\times&\left\{ \begin{array}{ll} K_0(m_B\sqrt{[\eta+(1-\eta)z](x_B+x_3-1)}b_3), \quad &x_B+x_3 \geq 1, \\
              \frac{i\pi}{2}H^{(1)}_0(m_B\sqrt{[\eta+(1-\eta)z](1-x_B-x_3)}b_3), \quad &x_B+x_3 < 1,\end{array} \right.  \nonumber\\
h^{\prime}_{2d}(x_B,z,x_3,b_B,b_3)&=&[K_0(m_B\sqrt{(1-\eta)z x_B}b_B)I_0(m_B\sqrt{(1-\eta)z x_B}b_3)\theta(b_B-b_3)+(b_3 \longleftrightarrow b_B) ] \nonumber\\
              &\times&\left\{ \begin{array}{ll} K_0(m_B\sqrt{r_c^2+(1-\eta)(x_B-x_3)z}b_3), \quad &r_c^2 \geq (1-\eta)(x_3-x_B)z,\\
              \frac{i\pi}{2}H^{(1)}_0(m_B\sqrt{(1-\eta)(x_3-x_B)z-r_c^2}b_3), \quad &r_c^2 < (1-\eta)(x_3-x_B)z,\end{array} \right. \nonumber\\
h_{2e}(z,x_3,b,b_3)&=&(\frac{i\pi}{2})^2 H^{(1)}_0(m_B\sqrt{(1-\eta)(1-z)x_3}b_3)
                [\theta(b_3-b)H^{(1)}_0(m_B\sqrt{1-(1-\eta)z-r_c^2}b_3) \nonumber\\
                &\times&  J_0(m_B\sqrt{1-(1-\eta)z-r_c^2}b)+(b_3 \longleftrightarrow b)]S_t(z),\nonumber\\
h_{2f}(z,x_3,b,b_3)&=&(\frac{i\pi}{2})^2 H^{(1)}_0(m_B\sqrt{(1-\eta)(1-z)x_3}b)[H^{(1)}_0(m_B\sqrt{x_3(1-\eta)}b) \nonumber\\
              &\times&J_0(m_B\sqrt{x_3(1-\eta)}b_3)\theta(b-b_3)+(b \longleftrightarrow b_3)]S_t(x_3), \nonumber\\
h_{2g}(x_B,z,x_3,b_B,b)&=&\frac{i\pi}{2} K_0(m_B\sqrt{1-(1-x_B-x_3) [\eta+z(1-\eta)]}b_B)[H^{(1)}_0(m_B\sqrt{(1-\eta)(1-z)x_3}b_B) \nonumber\\
              &\times&J_0(m_B\sqrt{(1-\eta)(1-z)x_3}b)\theta(b_B-b)+(b\longleftrightarrow b_B)], \nonumber\\
h_{2h}(x_B,z,x_3,b_B,b)&=&[\frac{i\pi}{2}H^{(1)}_0(m_B\sqrt{(1-\eta)(1-z)x_3}b_B)
              J_0(m_B\sqrt{(1-\eta)(1-z)x_3}b)\theta(b_B-b)+(b \longleftrightarrow b_B) ] \nonumber\\
              &\times&\left\{ \begin{array}{ll} K_0(m_B\sqrt{(1-\eta)(1-z)(x_B-x_3)}b_B), \quad &x_B \geq x_3, \\
              \frac{i\pi}{2}H^{(1)}_0(m_B\sqrt{(1-\eta)(1-z)(x_3-x_B)}b_B), \quad &x_B <  x_3,\end{array} \right.\nonumber
\end{eqnarray}
where $K_0$, $I_0$, and $H_0=J_0+iY_0$ are Bessel functions. The function $S_t(x)$ can be parametrized as
\begin{equation}
S_t(x)=\frac{2^{1+2c}\Gamma(3/2+c)}{\sqrt{\pi}\Gamma(1+c)}[x(1-x)]^c\;,\nonumber
\end{equation}
with $c=0.4$ for numerical calculation~\cite{Kurimoto:2001zj,Li:2009pr}.

\end{document}